\documentclass[10pt,conference]{IEEEtran}
\pdfoutput=1
\IEEEoverridecommandlockouts
\usepackage{cite}
\usepackage{amsmath,amssymb,amsfonts}
\usepackage{algorithmic}
\usepackage{graphicx}
\usepackage{textcomp}
\usepackage{xcolor}
\usepackage{url}
\usepackage{hyperref}
\usepackage{pifont}
\usepackage{array}
\newcommand{\cmark}{\text{\ding{51}}}
\newcommand{\xmark}{\text{\ding{55}}}
\usepackage[caption=false,font=footnotesize,labelfont=sf,textfont=sf]{subfig}
\usepackage{makecell}
\usepackage{balance}

\definecolor[named]{ACMBlue}{cmyk}{1,0.1,0,0.1}
\definecolor[named]{ACMYellow}{cmyk}{0,0.16,1,0}
\definecolor[named]{ACMOrange}{cmyk}{0,0.42,1,0.01}
\definecolor[named]{ACMRed}{cmyk}{0,0.90,0.86,0}
\definecolor[named]{ACMLightBlue}{cmyk}{0.49,0.01,0,0}
\definecolor[named]{ACMGreen}{cmyk}{0.20,0,1,0.19}
\definecolor[named]{ACMPurple}{cmyk}{0.55,1,0,0.15}
\definecolor[named]{ACMDarkBlue}{cmyk}{1,0.58,0,0.21}

\hypersetup{colorlinks=true,
  linkcolor=ACMRed,
  citecolor=ACMPurple,
  urlcolor=ACMDarkBlue,
  filecolor=ACMDarkBlue}

\def\BibTeX{{\rm B\kern-.05em{\sc i\kern-.025em b}\kern-.08em
    T\kern-.1667em\lower.7ex\hbox{E}\kern-.125emX}}
\begin{document}

\title{Tools and Benchmarks for Automated Log Parsing\thanks{\hspace{-2ex}\textsuperscript{*}This paper is accepted in ICSE'19.}}

\author{
\IEEEauthorblockN{Jieming Zhu\IEEEauthorrefmark{5}, Shilin He\IEEEauthorrefmark{2}, Jinyang Liu\IEEEauthorrefmark{3}, Pinjia He\IEEEauthorrefmark{4}, Qi Xie\IEEEauthorrefmark{6}, Zibin Zheng\IEEEauthorrefmark{3}, Michael R. Lyu\IEEEauthorrefmark{2}\vspace{2ex}}

\IEEEauthorblockA{
    \IEEEauthorrefmark{5}Huawei Noah's Ark Lab, Shenzhen, China\\
    \IEEEauthorrefmark{2}Department of Computer Science and Engineering, The Chinese University of Hong Kong, Hong Kong\\
    \IEEEauthorrefmark{3}School of Data and Computer Science, Sun Yat-Sen University, Guangzhou, China \\
    \IEEEauthorrefmark{4}Department of Computer Science, ETH Zurich, Switzerland\\
    \IEEEauthorrefmark{6}School of Computer Science and Technology, Southwest Minzu University, Chengdu, China\vspace{1ex}\\
    jmzhu@ieee.org,~~slhe@cse.cuhk.edu.hk,~~liujy@logpai.com,~~pinjiahe@gmail.com\\
    qi.xie.swun@gmail.com,~~zhzibin@mail.sysu.edu.cn,~~ lyu@cse.cuhk.edu.hk
}

}

\maketitle

\begin{abstract}
Logs are imperative in the development and maintenance process of many software systems. They record detailed runtime information that allows developers and support engineers to monitor their systems and dissect anomalous behaviors and errors. The increasing scale and complexity of modern software systems, however, make the volume of logs explodes. In many cases, the traditional way of manual log inspection becomes impractical. Many recent studies, as well as industrial tools, resort to powerful text search and machine learning-based analytics solutions. Due to the unstructured nature of logs, a first crucial step is to parse log messages into structured data for subsequent analysis. In recent years, automated log parsing has been widely studied in both academia and industry, producing a series of log parsers by different techniques. To better understand the characteristics of these log parsers, in this paper, we present a comprehensive evaluation study on automated log parsing and further release the tools and benchmarks for easy reuse. More specifically, we evaluate 13 log parsers on a total of 16 log datasets spanning distributed systems, supercomputers, operating systems, mobile systems, server applications, and standalone software. We report the benchmarking results in terms of accuracy, robustness, and efficiency, which are of practical importance when deploying automated log parsing in production. We also share the success stories and lessons learned in an industrial application at Huawei. We believe that our work could serve as the basis and provide valuable guidance to future research and deployment of automated log parsing. 
\end{abstract}


\section{Introduction}\label{sec:intro}

Logs play an important role in the development and maintenance of software systems. It is a common practice to record detailed system runtime information into logs, allowing developers and support engineers to understand system behaviours and track down problems that may arise. The rich information and the pervasiveness of logs enable a wide variety of system management and diagnostic tasks, such as analyzing usage statistics~\cite{Lee_vldb_2012}, ensuring application security~\cite{Oprea_dsn_2015}, identifying performance anomalies~\cite{Chow_osdi_2014,Nagaraj_nsdi_2012}, and diagnosing errors and crashes~\cite{Yuan_ASPLOSP10,XuDSN14}.

Despite the tremendous value buried in logs, how to analyze them effectively is still a great challenge~\cite{Oliner_cacm_2012}. First, modern software systems routinely generate tons of logs (e.g., about gigabytes of data per hour for a commercial cloud application~\cite{Mi_tpds_2013}). The huge volume of logs makes it impractical to manually inspect log messages for key diagnostic information, even provided with search and grep utilities. Second, log messages are inherently unstructured, because developers usually record system events using free text for convenience and flexibility~\cite{He_dsn_2016}. This further increases the difficulty in automated analysis of log data. Many recent studies (e.g., \cite{Du_ccs_2017, linqwICSE16, He_issre_2016}), as well as industrial solutions (e.g., Splunk~\cite{splunk}, ELK~\cite{elk}, Logentries~\cite{Logentries}), have evolved to provide powerful text search and machine learning-based analytics capabilities. To enable such log analysis, the first and foremost step is log parsing~\cite{He_dsn_2016}, a process to parse free-text raw log messages into a stream of structured events.



\begin{figure}[!t]
  \centering
  \includegraphics[width=0.48 \textwidth]{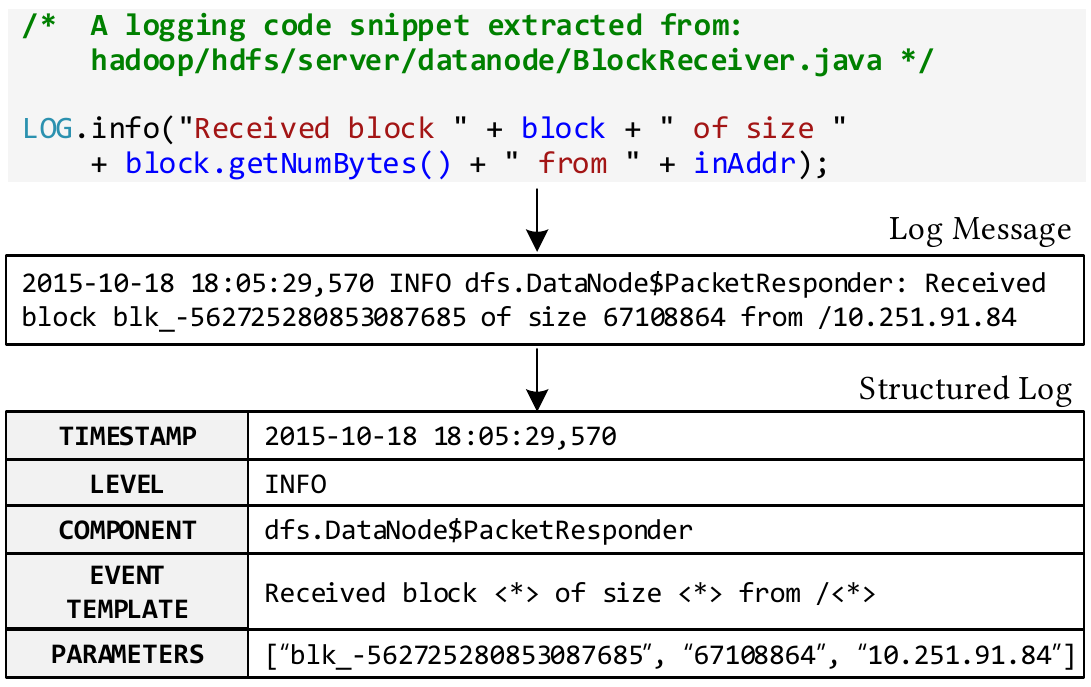}
  \caption{An Illustrative Example of Log Parsing}\label{fig:log_example}
\end{figure} 

As the example illustrated in Fig.\ref{fig:log_example}, each log message is printed by a logging statement and records a specific system event with its message header and message content. The message header is determined by the logging framework and thus can be relatively easily extracted, such as timestamp, verbosity level (e.g., ERROR/INFO/DEBUG), and component. In contrast, it is often difficult to structurize the free-text message content written by developers, since it is a composition of \texttt{constant} strings and \texttt{variable} values. The constant part reveals the event template of a log message and remains the same for every event occurrence. The variable part carries dynamic runtime information (i.e., parameters) of interest, which may vary among different event occurrences. The goal of log parsing is to convert each log message into a specific event template (e.g., ``\texttt{Received block} \verb|<|*\verb|>| \texttt{of size} \verb|<|*\verb|>| \texttt{from} /\verb|<|*\verb|>|") associated with key parameters (e.g., [``\texttt{blk\_-562725280853087685}", ``\texttt{67108864}", ``\texttt{10.251.91.84}"]). Here, ``\verb|<|*\verb|>|" denotes the position of each parameter.  


The traditional way of log parsing relies on handcrafted regular expressions or grok patterns~\cite{Grok} to extract event templates and key parameters. Although straightforward, manually writing ad-hoc rules to parse a huge volume of logs is really a time-consuming and error-prone pain  (e.g., over 76K templates in our Android dataset). Especially, logging code in modern software systems usually update frequently (up to thousands of log statements every month~\cite{XuPhd}), leading to the inevitable cost of regularly revising these handcrafted parsing rules. To reduce the manual efforts in log parsing, some studies~\cite{Xu_sosp_2009,Nagappan_issre_2009} have explored the static analysis techniques to extract event templates from source code directly. While it is a viable approach in some cases, source code is not always accessible in practice (e.g., when using third-party components). Meanwhile, non-trivial efforts are required to build such a static analysis tool for software systems developed across different programming languages. 

To achieve the goal of automated log parsing, many data-driven approaches have been proposed from both academia and industry, including frequent pattern mining (SLCT~\cite{SLCT03}, and its extension LogCluster~\cite{logcluster15}), iterative partitioning (IPLoM~\cite{IPLoM09}), hierarchical clustering (LKE~\cite{fu2009execution}), longest common subsequence computation (Spell~\cite{spell16}), parsing tree (Drain~\cite{drain17}), etc. In contrast to handcrafted rules and source code-based parsing, these approaches are capable of learning patterns from log data and automatically generating common event templates. In our previous work~\cite{He_dsn_2016}, we have conducted an evaluation study of four representative log parsers and made the first step towards reproducible research and open-source tools for automated log parsing. This, to some extent, facilitates some recent developments of tools such as LenMa~\cite{lenma16}, LogMine~\cite{logmine16}, Spell~\cite{spell16}, Drain~\cite{drain17}, and MoLFI~\cite{molfi18}. Even more, automated log parsing lately becomes an appealing selling point in some trending log management solutions (e.g., Logentries~\cite{Logentries} and Loggly~\cite{Loggly}). 

In this paper, we present a more comprehensive study on automated log parsing and further publish a full set of tools and benchmarks to researchers and practitioners. In reality, companies are usually reluctant to open their system logs due to confidential issues, leading to the scarcity of real-world log data. 
With close collaborations with our industrial partners, as well as some pioneer researchers (authors from~\cite{Du_ccs_2017,BGLdata,Xu_sosp_2009}), we collect a large set of logs (over 77GB in total) produced by 16 different systems spanning distributed systems, supercomputers, operating systems, mobile systems, server applications, and standalone software. Since the first release of these logs~\cite{loghub}, they have been requested by over 150 organizations from both industry and academia.

Meanwhile, the lack of publicly-available
tools hinders the adoption of automated log parsing. Therefore, we release an easy-to-use, open-source toolkit\footnote{\url{https://github.com/logpai/logparser}}, with a total of 13 recently-published log parsing methods. We evaluate them thoroughly on 16 different log datasets and report the results in terms of accuracy, robustness, and efficiency. The benchmarking results could help users better understand the characteristics of different log parsers and guide the deployment of automated log parsing in production. We also share the success stories and lessons learned in an industrial application at Huawei. We believe that the availability of tools and benchmarks, as well as the industrial experiences shared in this study, would benefit future research and facilitate wide adoption of automated log parsing in industry.





The remainder of the paper is organized as follows. Section~\ref{sec:background} reviews the state-of-the-art log parsers. Section~\ref{sec:experiment} reports the benchmarking results. We share our industrial deployment in Section~\ref{sec:case_study}, and summarize the related work in Section~\ref{sec:relatedwork}. Finally, we conclude the paper in Section~\ref{sec:conclusion}.

\section{Log Parsing}\label{sec:background}

\begin{table*}[!ht]
\centering
\caption{Summary of Industrial Log Management Tools and Services} \label{tab:industryparsers}
\includegraphics[width=1\textwidth]{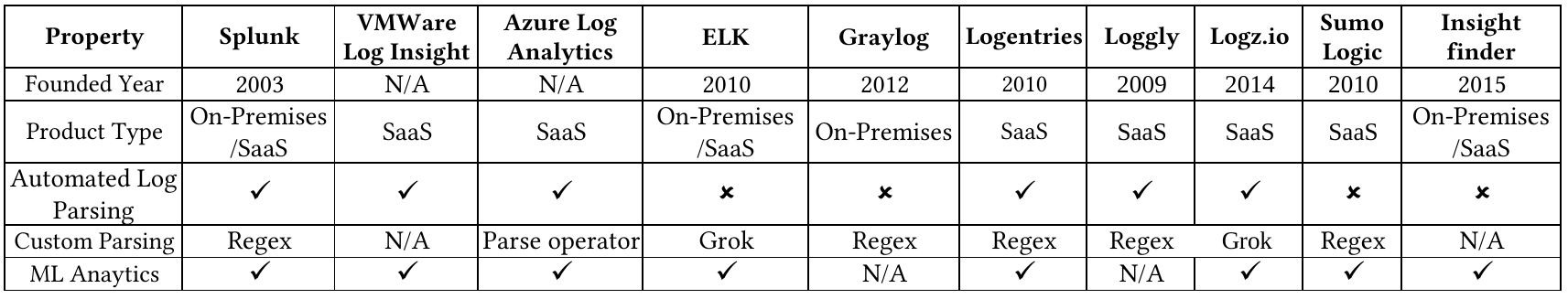}
\vspace{-2em}
\end{table*} 

\begin{table*}[!ht]
\centering
\caption{Summary of Automated Log Parsing Tools} \label{tab:logparsers}
\includegraphics[width=0.86\textwidth]{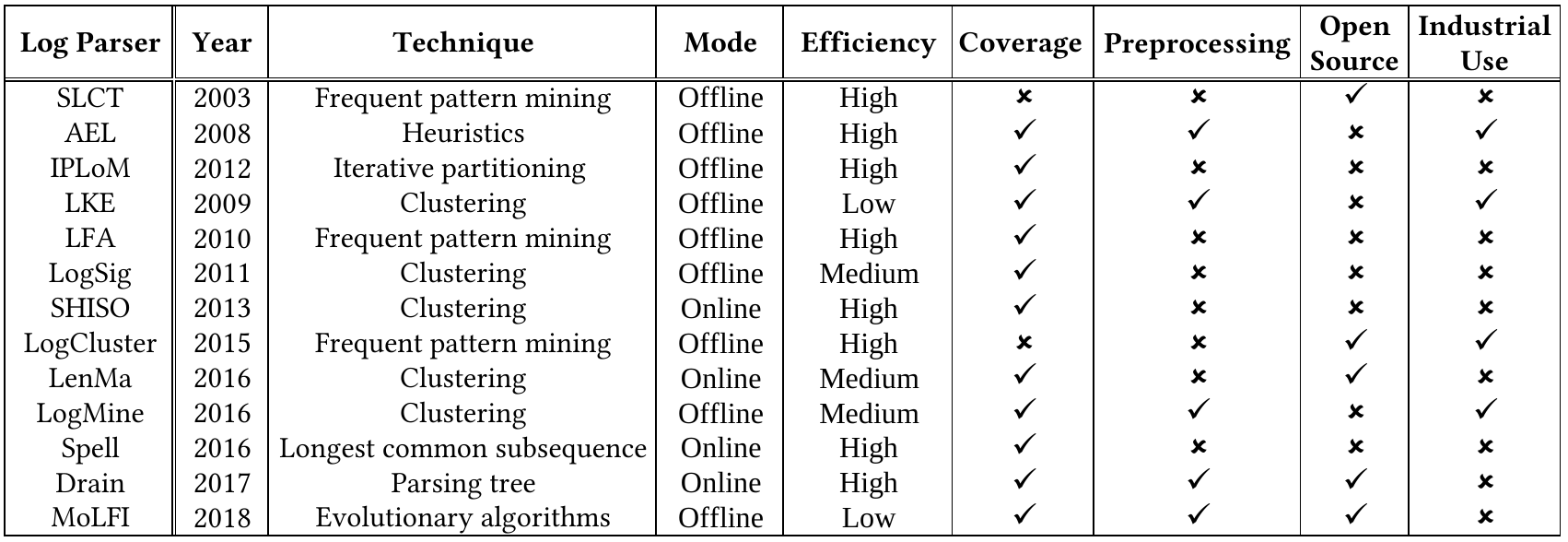}
\vspace{-2em}
\end{table*} 

In this section, we present some motivating applications of log parsing, review the characteristics and techniques of existing log parsers, and then describe our tool implementation.

\subsection{Motivating Applications}
Log parsing typically serves as the first step towards downstream log analysis tasks. Parsing textual log messages into a structured format enables efficient search, filtering, grouping, counting, and sophisticated mining of logs. To illustrate, we provide a list of sample industrial applications here, which have been widely studied by researchers and practitioners. 
\begin{itemize}
    \item \textit{Usage analysis}. Employing logs for usage analysis is a common task during software development and maintenance. Typical examples include user behaviour analysis (e.g., Twitter~\cite{Lee_vldb_2012}), API profiling, log-based metrics counting (e.g., Google Cloud~\cite{google_cloud}), and workload modeling (e.g., Microsoft~\cite{Magpie}). These applications typically require structured events as inputs. 
    \item \textit{Anomaly detection}. Anomaly detection nowadays plays a central role in system monitoring. Logs record detailed execution information and thus serve as a valuable data source to detect abnormal system behaviours. Some recent work has investigated the use of machine learning techniques (e.g., PCA~\cite{Xu_sosp_2009}, invariant mining~\cite{jglouATC10}, and deep learning~\cite{Du_ccs_2017}) for anomaly detection. In such cases, log parsing is a necessary data preprocessing step to train machine learning models.
    \item \textit{Duplicate issue identification}. In practice, system issues (e.g., disk error, network disconnection) often recur or can be repeatedly reported by different users, leading to many duplicate issues. It is crucial to automatically identify duplicate issues to reduce the efforts of developers and support engineers. Microsoft has reported some studies~\cite{ding2014mining,Lim_icdm_14,linqwICSE16} on this task, in which structured event data are required.
    \item \textit{Performance modeling}. Facebook has recently reported a use case~\cite{Chow_osdi_2014} to apply logs as a valuable data source to performance modeling, where potential performance improvements can be quickly validated. A prerequisite to this approach is to extract all possible event templates from the logs. The performance model construction takes event sequences as inputs. 
    \item \textit{Failure diagnosis}. Manual failure diagnosis is a time consuming and challenging task since logs are not only of huge volume but also extremely verbose and messy. Some recent progress~\cite{Nagaraj_nsdi_2012,Oreilly17} has been made to automate root cause analysis based on machine learning techniques. Likewise, log parsing is deemed as a prerequisite.
\end{itemize}

\subsection{Characteristics of Log Parsers}
As an important step in log analysis, automated approaches of log parsing have been widely studied, producing an abundance of log parsers ranging from research prototypes to industrial solutions. To gain an overview of existing log parsers, we summarize the key characteristics of them.

\textbf{1) Industrial Solutions}. 
Table~\ref{tab:industryparsers} provides a summary of some industrial log analysis and management tools. With the upsurge of big data, many cloud providers as well as startup companies provide on-premise or software-as-a-service (SaaS) solutions for log management. They enable powerful log search, visualization, and machine learning (ML) analytics capabilities. To illustrate, we list 10 representative products in the market, including both well-established ones (e.g., Splunk~\cite{splunk}) and newly-started ones (e.g., Logz.io~\cite{logz}). As a key component, automated log parsing has recently risen as a appealing selling point in some products~\cite{rapid_autoparsing,logz_autoparsing,loggly_autoparsing}. Current solutions of automated log parsing, however, are realized with built-in parsing support for common log types, such as Apache and Nginx logs~\cite{rapid_autoparsing}. For other types of logs, they have to rely on users to perform custom parsing with regex scripts, grok patterns~\cite{Grok}, or a parsing wizard. Current industrial parsing solutions require deep domain knowledge, and thus fall out of the scope of this study.

\textbf{2) Research Studies}.
Table~\ref{tab:logparsers} provides a summary of 13 representative log parsers proposed in the literature, which are the main subjects of our study. These log parsers are all aimed for automated log parsing, but may differ in quality. After reviewing the literature, we list some key characteristics for log parsers that are of practical importance.

\textbf{Technique.} Different log parsers may adopt different log parsing strategies. We categorize them into 7 types of strategies, including frequent pattern mining, clustering, iterative partitioning, longest common subsequence, parsing tree, evolutionary algorithms, and other heuristics. We will present more details of these log parsing methods in Section \ref{sec:parsers}. 

\textbf{Mode.} According to different scenarios of log parsing, log parsers can be categorized to two main modes, i.e., offline and online. Offline log parsers are a type of batch processing and require that all the log data are available before parsing. On the contrary, online log parsers process log messages one by one in a streaming manner, which is often more practical when logs are collected as a stream.

\textbf{Efficiency.} Efficiency is always a major concern for log parsing in practice, considering the large volume of logs. An inefficient log parser can greatly hinder subsequent log analysis tasks that have low latency requirements in cases such as real-time anomaly detection and performance monitoring. In Table \ref{tab:logparsers}, the efficiency of current tools has been categorized into three levels: high, medium and low. 

\textbf{Coverage.} Coverage denotes the capability of a log parser to successfully parse all input log messages. If yes, it is marked as ``\cmark". ``\xmark" indicates that a log parser can only structurize part of the logs. For example, SLCT can extract frequently-occurring event templates by applying frequent pattern mining, but fails to handle rare event templates precisely. A high-quality log parser should be able to process all input log messages, since ignoring any important event may miss the opportunity for anomaly detection and root cause identification.

\textbf{Preprocessing.} Preprocessing is a step to remove some common variable values, such as IP address and numbers, by manually specifying simple regular expressions. The preprocessing step is straightforward, but require some additional manual work. We mark ``\cmark" if a preprocessing step is explicitly specified in a log parsing method, and ``\xmark" otherwise.



\textbf{Open-source.} An open-source log parser can allow researchers and practitioners to easily reuse and further improve existing log parsing methods. This can not only benefit related research but also facilitate wide adoption of automated log parsing. However, current open-source tools for log parsing are still limited. We mark ``\cmark" if an existing log parser is open-source, and ``\xmark" otherwise. 

\textbf{Industrial use.}  A log parser has more practical value and should be more reliable if it has been deployed in production for industrial use. We mark ``\cmark" if a log parser has been reported on use in an industrial setting, and ``\xmark" otherwise.

\subsection{Techniques of Log Parsers}\label{sec:parsers}
In this work, we have studied a total of 13 log parsers. We briefly summarize the techniques used by these log parsers from the following aspects:

\subsubsection{Frequent Pattern Mining} A frequent pattern is a set of items that occurs frequently in a data set. Likewise, event templates can be seen as a set of constant tokens that occurs frequently in logs. Therefore, frequent pattern mining is an straightforward approach to automated log parsing. Examples include SLCT \cite{SLCT03}, LFA \cite{LFA_10}, and LogCluster \cite{logcluster15}. All the three log parsers are offline methods and follow a similar parsing procedure: 1) traversing over the log data by several passes, 2) building frequent itemsets (e.g., tokens, token-position pairs) at each traversal, 3) grouping log messages into several clusters, and 4) extracting event templates from each cluster. SLCT, to our knowledge, is the first work that applies frequent pattern mining to log parsing. Furthermore, LFA considers the token frequency distribution in each log message instead of the whole log data to parse rare log messages. LogCluster is an extension of SCLT, and can be robust to shifts in token positions.

\subsubsection{Clustering} Event template forms a natural pattern of a group of log messages. From this view, log parsing can be modeled as a clustering problem of log messages. Examples that apply the clustering algorithms for log parsing include 3 offline methods (i.e., LKE \cite{fu2009execution}, LogSig \cite{tang2011logsig}, and LogMine \cite{logmine16}) and 2 online methods (i.e., SHISO \cite{SHISO_13}, and LenMa \cite{lenma16}). Specifically, LKE employs the hierarchical clustering algorithm based on weighted edit distances between pairwise log messages. LogSig is a message signature based algorithm to cluster log messages into a predefined number of clusters. LogMine can generate event templates in a hierarchical clustering way, which groups log messages into clusters from bottom to top. SHISO and LenMa are both online methods, which parse logs in a similar streaming manner. For each newly coming log message, the parsers first compute its similarity to representative event templates of existing log clusters. The log message will be added to an existing cluster if it is successfully matched, otherwise a new log cluster will be created. Then, the corresponding event template will be updated accordingly. 

\subsubsection{Heuristics} Different from general text data, log messages have some unique characteristics. As such, some work (i.e., AEL \cite{AEL_1}, IPLoM \cite{IPLoM09}, Drain \cite{drain17}) proposes heuristics-based log parsing methods. Specifically, AEL separates log messages into multiple groups by comparing the occurrences between constant tokens and variable tokens. IPLoM employs an iterative partitioning strategy, which partitions log messages into groups by message length, token position and mapping relation. Drain applies a fixed-depth tree structure to represent log messages and extracts common templates efficiently. These heuristics make use of the characteristics of logs and perform quite well in many cases.

\subsubsection{Others} Some other methods exist. For example, Spell \cite{spell16} utilizes the longest common subsequence algorithm to parse logs in a stream manner. Recently, Messaoudi et al. \cite{molfi18} propose MoLFI, which models log parsing as a multiple-objective optimization problem and solves it using evolutionary algorithms. 

\begin{table*}[!tbp]
\centering \caption{Summary of Loghub Datasets} \label{tab:datasets}
\setcellgapes{0.03in}
\makegapedcells 
\begin{tabular}{|c||l|r|r|r|c|c|}
\hline
{\textbf{Dataset}}&{\textbf{Description}}&\textbf{Time Span}& \textbf{Data Size}& \textbf{\#Messages}& \textbf{\#Templates (total)} & \textbf{\#Templates (2k)}\\
\hline
\multicolumn{7}{|c|}{Distributed system logs} \\
\hline
HDFS & Hadoop distributed file system log & 38.7 hours & 1.47 GB & 11,175,629 & 30 & 14\\
\hline
Hadoop & Hadoop mapreduce job log & N.A. & 48.61 MB  & 394,308 & 298 & 114\\
\hline
Spark & Spark job log & N.A. & 2.75 GB & 33,236,604  & 456 & 36\\
\hline
ZooKeeper & ZooKeeper service log & 26.7 days & 9.95 MB & 74,380  & 95 & 50\\
\hline
OpenStack & OpenStack software log & N.A. & 60.01 MB & 207,820  & 51 & 43\\
\hline
\multicolumn{7}{|c|}{Supercomputer logs} \\
\hline
BGL & Blue Gene/L supercomputer log & 214.7 days & 708.76 MB & 4,747,963 & 619 &120\\
\hline
HPC & High performance cluster log & N.A. & 32.00 MB & 433,489 & 104 & 46\\
\hline
Thunderbird & Thunderbird supercomputer log & 244 days & 29.60 GB & 211,212,192 & 4,040 &149\\
\hline
\multicolumn{7}{|c|}{Operating system logs} \\
\hline
Windows & Windows event log & 226.7 days & 26.09 GB & 114,608,388 & 4,833 &50 \\
\hline
Linux & Linux system log & 263.9 days & 2.25 MB & 25,567 & 488 & 118\\
\hline
Mac & Mac OS log & 7.0 days & 16.09 MB & 117,283 & 2,214 & 341\\
\hline
\multicolumn{7}{|c|}{Mobile system logs} \\
\hline
Android & Android framework log & N.A. & 3.38 GB & 30,348,042 & 76,923 & 166\\
\hline
HealthApp & Health app log & 10.5 days & 22.44 MB & 253,395 & 220& 75\\
\hline
\multicolumn{7}{|c|}{Server application logs} \\
\hline
Apache & Apache server error log & 263.9 days & 4.90 MB & 56,481  & 44 & 6\\
\hline
OpenSSH & OpenSSH server log & 28.4 days & 70.02 MB & 655,146 & 62 & 27\\
\hline
\multicolumn{7}{|c|}{Standalone software logs} \\
\hline
Proxifier & Proxifier software log & N.A. & 2.42 MB & 21,329 & 9 & 8\\
\hline
\end{tabular}
\vspace{-2em}
\end{table*}

\subsection{Tool Implementation}
Although automated log parsing has been studied for several years, it is still not a well-received technique in industry. This is largely due to the lack of publicly available tools that are ready for industrial use. For operation engineers who often have limited expertise in machine learning techniques, implementing an automated log parsing tool requires non-trivial efforts. This may exceed the overhead for manually crafting regular expressions. Our work aims to bridge this gap between academia and industry and promote the adoption for automated log parsing. We have implemented an open-source log parsing toolkit, namely logparser, and released a large benchmark set as well. As a part-time project, the implementation of logparser takes over two years and have 11.7K LOC in Python. Currently, logparser contains a total of 13 log parsing methods proposed by researchers and practitioners. Among them, five log parsers (i.e., SLCT, LogCluster, LenMa, Drain, MoLFI) are open-source from existing research work. However, they are implemented in different programming languages and have different input/output formats. Examples and documents are also missing or incomplete, making it difficult for a trial. For ease of use, we define a standard and unified input/output interface for different log parsing methods and further wrap up the existing tools into a single Python package. Logparser requires a raw log file with free-text log messages as input, and finally outputs a structured log file and an event template file with aggregated event counts. The outputs can be easily fed into subsequent log mining tasks. Our logparser toolkit can help engineers quickly identify the strengths and weaknesses of different log parsing methods and evaluate their possibility for industrial use cases.


\section{Evaluation}\label{sec:experiment}

\begin{table*}[!ht]
\centering
\caption{Accuracy of Log Parsers on Different Datasets} \label{tab:accu_comparison}
\includegraphics[width=1\textwidth]{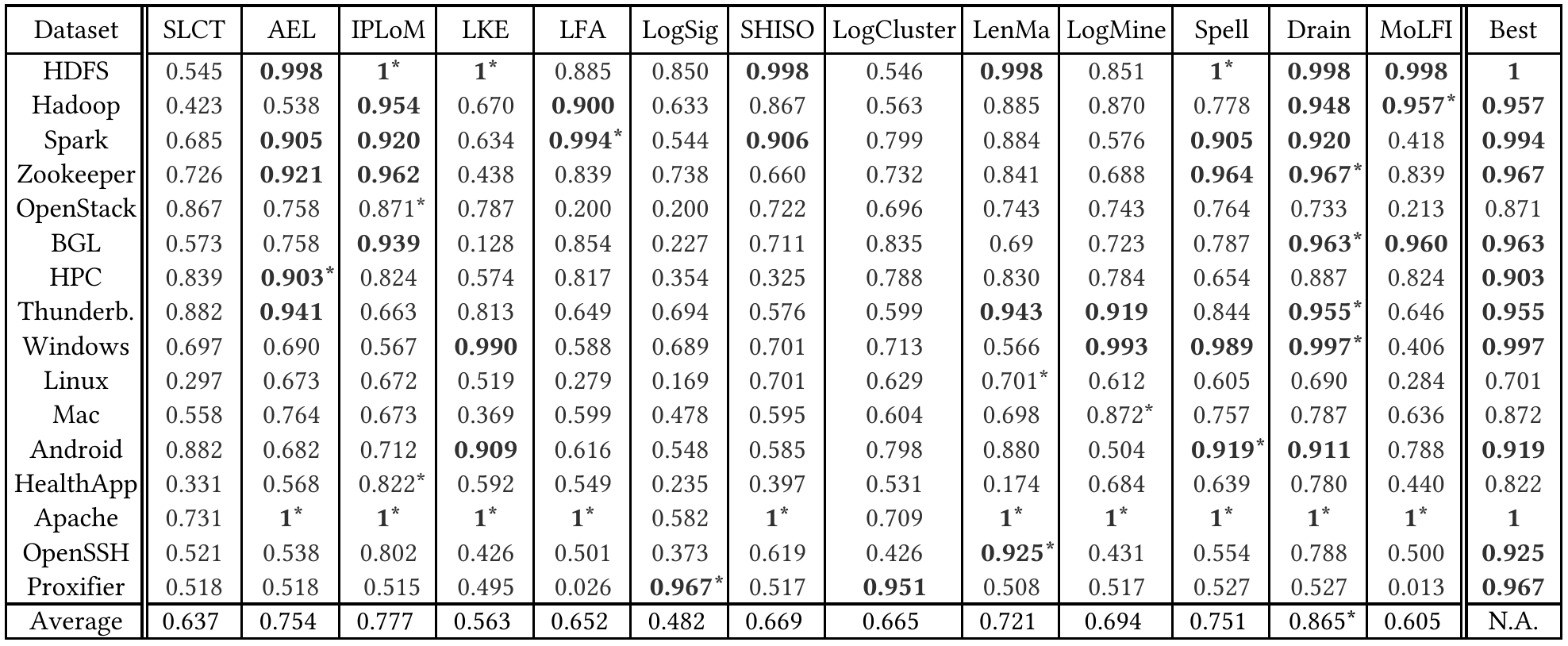}
\vspace{-3em}
\end{table*}

In this section, we evaluate 13 log parsers on 16 benchmark datasets, and report the benchmarking results in terms of accuracy, robustness, and efficiency. They are three key qualities of interest when applying log parsing in production. 
\begin{itemize}
    \item \textit{Accuracy} measures the ability of a log parser in distinguishing constant parts and variable parts. Accuracy is one main focus of existing log parsing studies, because an inaccurate log parser could greatly limit the effectiveness of the downstream log mining tasks \cite{He_dsn_2016}.
    \item  \textit{Robustness} of a log parser measures the consistency of its accuracy under log datasets of different sizes or from different systems. A robust log parser should perform consistently across different datasets, and thus can be used in the versatile production environment.
    \item  \textit{Efficiency} measures the processing speed of a log parser. We evaluate the efficiency by recording the time that a parser takes to parse a specific dataset. The less time a log parser consumes, the higher efficiency it provides.
\end{itemize}

\subsection{Experimental Setup}
\textbf{Dataset}. Real-world log data are currently scarce in public due to confidential issues, which hinders the research and development of new log analysis techniques. In this work, we have released, on our loghub data repository~\cite{loghub}, a large collection of logs from 16 different systems spanning distributed systems, supercomputers, operating systems, mobile systems, server applications, and standalone software. Table~\ref{tab:datasets} presents a summary of the datasets. Some of them (e.g., HDFS~\cite{Xu_sosp_2009}, Hadoop~\cite{linqwICSE16}, BGL~\cite{BGLdata}) are production logs released by previous studies, while the others (e.g., Spark, Zookeeper, HealthApp, Android) are collected from real-world systems in our lab. Loghub contains a total of 440 million log messages that amounts to 77 GB in size. To the best of our knowledge, it is the largest collection of log datasets. Wherever possible, the logs are not sanitized, anonymized or modified in any way. They are freely accessible for research purposes. At the time of writing, our loghub datasets have been downloaded over 1000 times by more than 150 organizations from both industry (35\%) and academia (65\%).

In this work, we use the loghub datasets as benchmarks to evaluate all existing log parsers. The large size and diversity of loghub datasets can not only measure the accuracy of log parsers but also test the robustness and efficiency of them. To allow easy reproduction of the benchmarking results, we randomly sample 2000 log messages from each dataset and manually label the event templates as ground truth. Specifically, in Table~\ref{tab:datasets}, ``\#Templates (2k sample)" indicates the number of event templates in log samples, while ``\#Templates (total)" shows the total number of event templates generated by a rule-based log parser.

\textbf{Accuracy Metric}. To quantify the effectiveness of automated log parsing, as with~\cite{spell16}, we define the parsing accuracy (PA) metric as the ratio of correctly parsed log messages over the total number of log messages. After parsing, each log message has an event template, which in turn corresponds to a group of messages of the same template. A log message is considered correctly parsed if and only if its event template corresponds to the same group of log messages as the ground truth does. For example, if a log sequence [E1, E2, E2] is parsed to [E1, E4, E5], we get PA=$1/3$, since the 2nd and 3rd messages are not grouped together. In contrast to standard evaluation metrics that are used in previous studies, such as precision, recall, and F1-measure~\cite{IPLoM09,He_dsn_2016,molfi18}, PA is a more rigorous metric. In PA, partially matched events are considered incorrect.




For fairness of comparison, we apply the same preprocessing rules (e.g., IP or number replacement) to each log parser. The parameters of all the log parsers are fine tuned through over 10 runs and the best results are reported to avoid bias from randomization. All the experiments were conducted on a server with 32 Intel(R) Xeon(R) 2.60GHz CPUs, 62GB RAM, and Ubuntu 16.04.3 LTS installed.


\subsection{Accuracy of Log Parsers}
\begin{figure*}[!t]
  \centering
  \includegraphics[width=0.85 \textwidth]{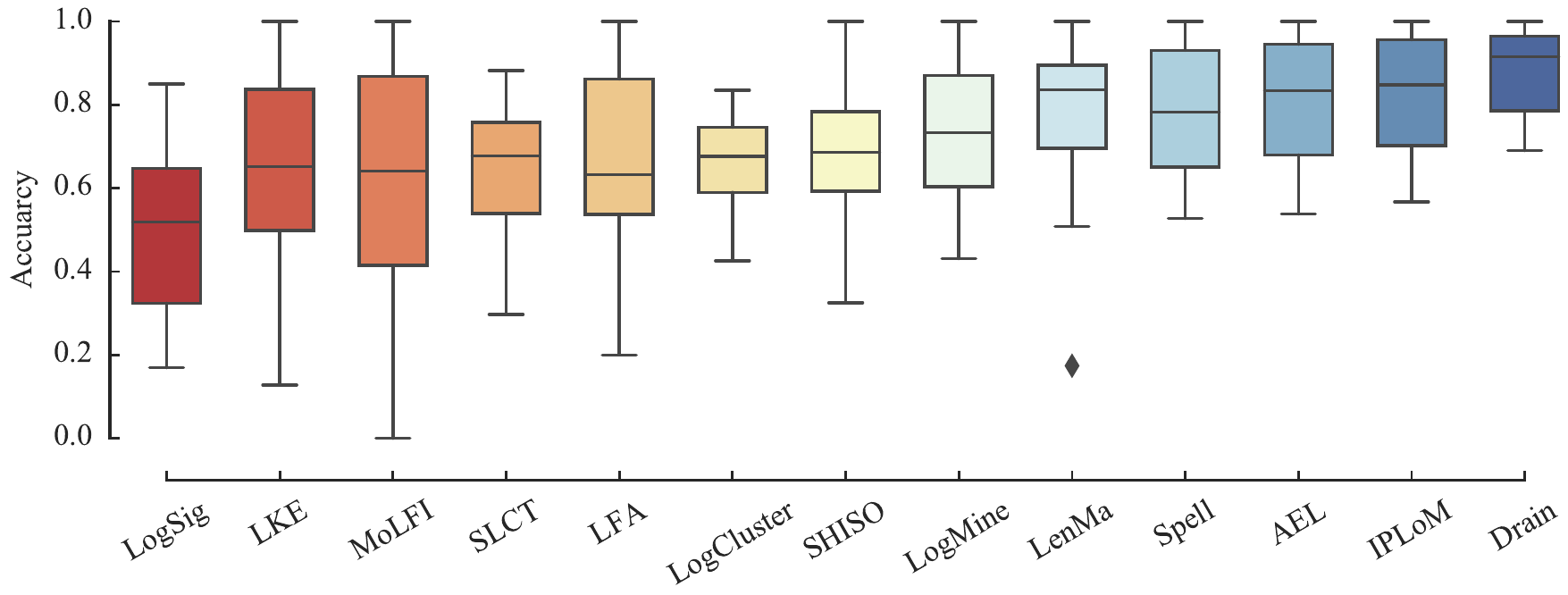}
  \vspace{-1em}
  \caption{Accuracy Distribution of Log Parsers across Different Types of Logs}\label{fig:accu_boxplot}
  \vspace{-1em}
\end{figure*}
%

In this part, we evaluate the accuracy of log parsers. We found that some log parsers (e.g., LKE) cannot handle the original datasets in reasonable time (e.g., even days). Thus, for fair comparison, the accuracy experiments are conducted on sampled subsets, each containing 2,000 log messages. The log messages are randomly sampled from the original log dataset, yet retains the key properties, such as event redundancy and event variety.

Table \ref{tab:accu_comparison} presents the accuracy results of 13 log parsers evaluated on 16 log datasets. Specifically, each row denotes the parsing accuracy of different log parsers on one dataset, which facilitates comparison among different log parsers. Each column represents the parsing accuracy of one log parser over different datasets, which helps identify its robustness across different types of logs. In particular, we mark accuracy values greater than 0.9 in boldface since they indicate high accuracy in practice. For each dataset, the best accuracy is highlighted with a asterisk ``*" and shown in the column ``Best". We can observe that most of the datasets are accurately (over $90\%$) parsed by at least one log parser. Totally, 8 out of 13 log parsers attatin the best accuracy on at least two log datasets. Even more, some log parsers can parse the HDFS and Apache datasets with $100\%$ accuracy. This is because HDFS and Apache error logs have relatively simple event templates and are easy to identify. However, several types of logs (e.g., OpenStack, Linux, Mac, HealthApp) still could not be parsed accurately due to their complex structure and abundant event templates (e.g., 341 templates in Mac logs). Therefore, further improvements should be made towards better parsing those complex log data. 

To measure the overall effectiveness of log parsers, we compute the average accuracy of each log parser across different datasets, as shown in the last row of Table \ref{tab:accu_comparison}. We can observe that, on average, the most accurate log parser is Drain, which attains high accuracy on 9 out of 16 datasets. The other top ranked log parsers include IPLoM, AEL, and Spell, which achieve high accuracy on 6 datasets. In contrast, the four log parsers that have the lowest average accuracy are LogSig, LFA, MoLFI, and LKE. From the results, we can briefly conclude that log parsers should take full advantage of the inherent structure and characteristics of log messages to achieve good parsing accuracy, instead of directly applying standard algorithms such as clustering and frequent pattern mining.

\subsection{ Robustness of Log Parsers}

Robustness is crucial to the practical use of a log parser in production environments. In this part, we evaluate the robustness of log parsers from two aspects: 1) robustness across different types of logs and 2) robustness on different volumes of logs. 

Figure \ref{fig:accu_boxplot} shows a boxplot that indicates the accuracy distribution of each log parser across the 16 log datasets. For each box, the horizontal lines from bottom to top correspond to the minimum, 25-percentile, median, 75-percentile and maximum accuracy values. The diamond mark denotes an outlier point, since LenMa only has an accuracy of 0.174 on HealthApp logs. From left to right in the figure, the log parsers are arranged in ascending order of the average accuracy shown in Table~\ref{tab:accu_comparison}. That is, LogSig has the lowest accuracy and Drain obtains the highest accuracy on average. A good log parser should be able to parse many different types of logs for general use. However, we can observe that, although most log parsers achieve the maximal accuracy over 0.9, they have a large variance over different datasets. There is still no log parser that performs well on all log data. Therefore, we suggest users to try different log parsers on their own logs first. Currently, Drain performs the best among all the 13 log parsers under study. It not only attains the highest accuracy on average, but also shows the smallest variance. 

\begin{figure*}[!t]
\begin{minipage}[b]{1\linewidth}
    \centering
    \subfloat[HDFS]{
        \includegraphics[width=0.32
        \textwidth]{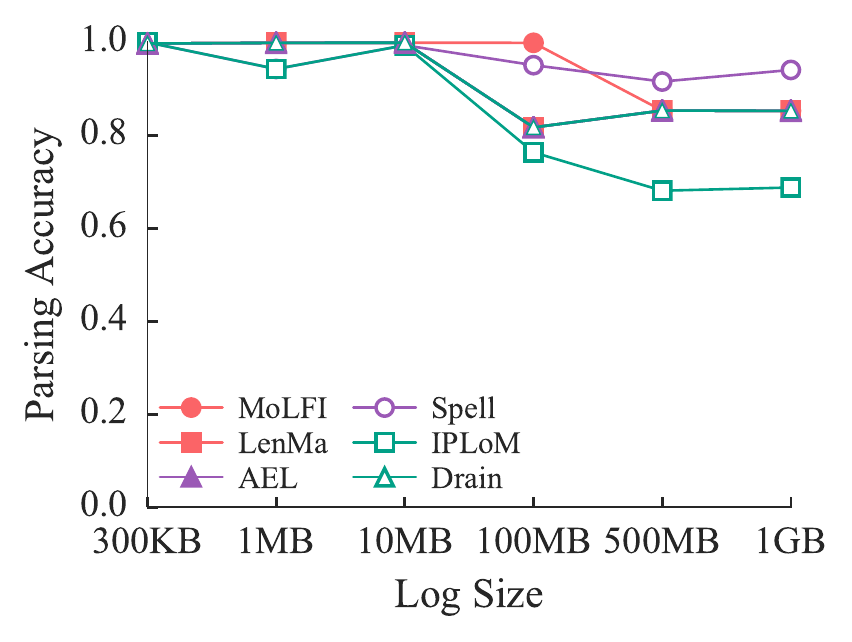}}
        \hfil
    \subfloat[BGL]{
        \includegraphics[width=0.32
        \textwidth]{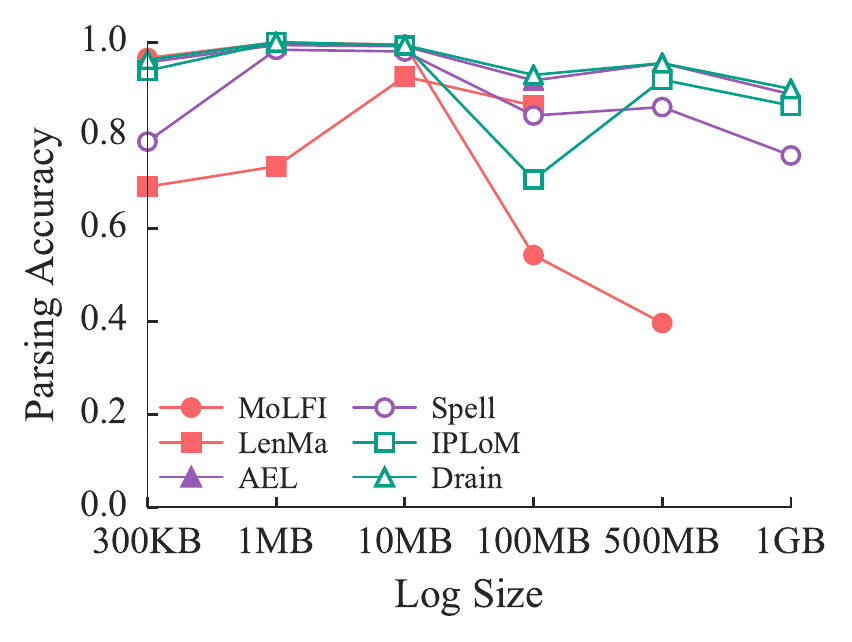}}
        \hfil
    \subfloat[Android]{
        \includegraphics[width=0.32
        \textwidth]{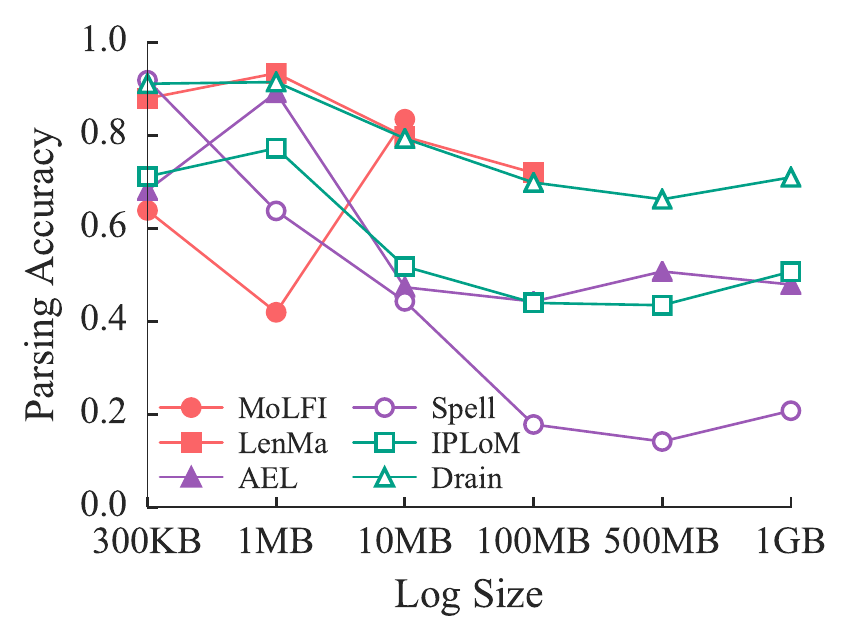}}
    \caption{Accuracy of Log Parsers on Different Volumes of Logs}\label{fig:accuracy_datasize}
\end{minipage}
\vspace{-5ex}
\end{figure*}

\begin{figure*}[!t]
\begin{minipage}[b]{1\linewidth}
    \centering
    \subfloat[HDFS]{
        \includegraphics[width=0.32
        \textwidth]{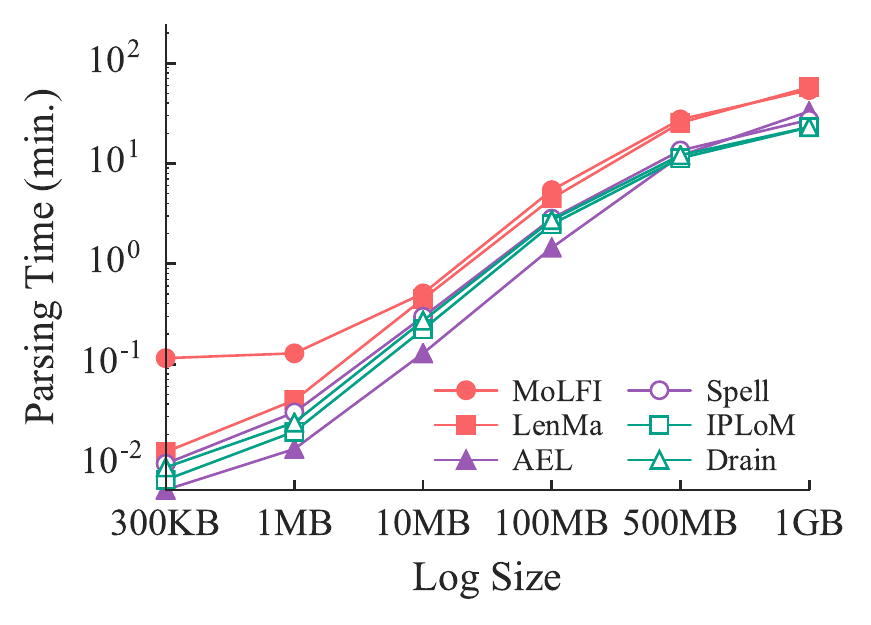}}
        \hfil
    \subfloat[BGL]{
        \includegraphics[width=0.32
        \textwidth]{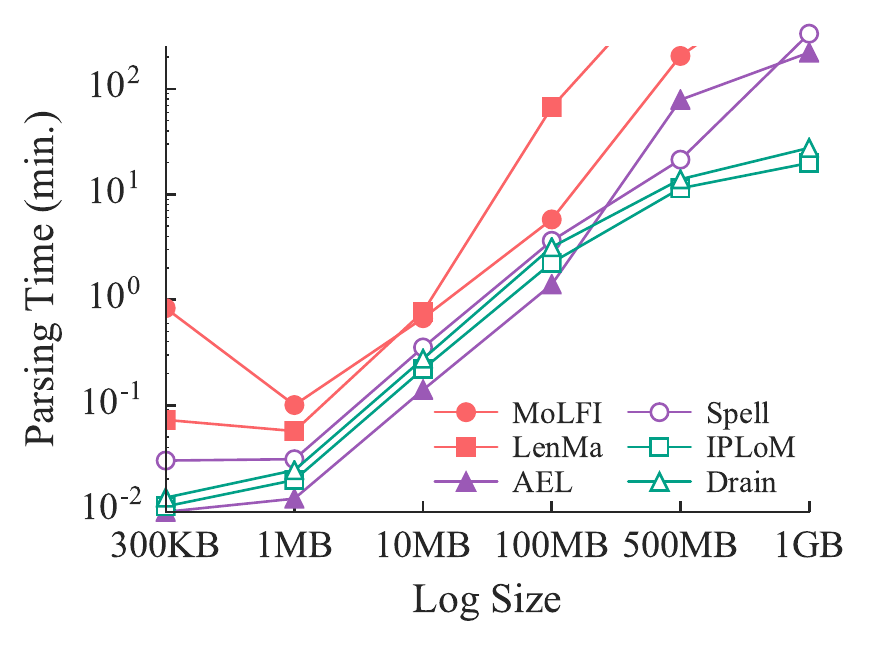}}
		\hfil
    \subfloat[Android]{
        \includegraphics[width=0.32
        \textwidth]{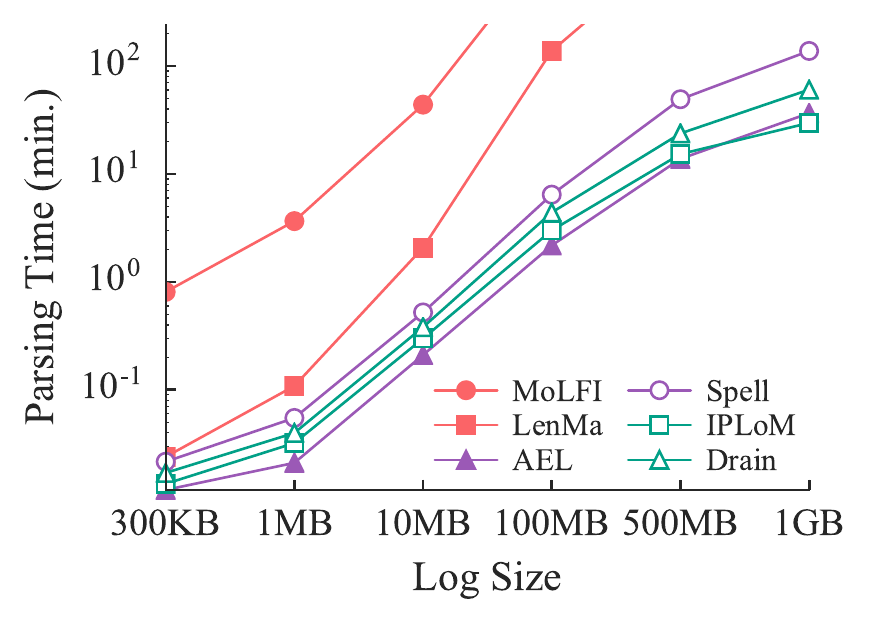}}
    \caption{Efficiency of Log Parsers on Different Volumes of Logs}\label{fig:time_datasize}
\end{minipage}
\vspace{-5ex}
\end{figure*}

In addition, we evaluate the robustness of log parsers on different volumes of logs. In this experiment, we select six log parsers, i.e., MoLFI, Spell, LenMa, IPLoM, AEL, and Drain. They have achieved high accuracy (over $90\%$) on more than four log datasets, as shown in Table~\ref{tab:accu_comparison}. Meanwhile, MoLFI is the most recently published log parser, and the other five log parsers are ranked in the top in Figure \ref{fig:accu_boxplot}. We also choose three large datasets, i.e., HDFS, BGL, and Android. The raw logs have a volume of over 1GB each, and the groundtruth templates are readily available for accuracy computation. HDFS and BGL have also been used as benchmarks datasets in the previous work~\cite{spell16, IPLoM09}. For each log dataset, we vary the volume from 300 KB to 1 GB, while fix the parameters of log parsers that were fine tuned on 2k log samples. Specifically, 300KB is roughly the size of each 2k log sample. We truncate the raw log files to obtain samples of other volumes (e.g., 1GB). Figure \ref{fig:accuracy_datasize} shows the parsing accuracy results. Note that some lines are incomplete in the figure, because methods like MoLFI and LenMa cannot finish parsing within reasonable time (6 hours in our experiment). A good log parser should be robust to such changes of log volumes. However, we can see that parameters tuned on small log samples cannot fit well to large log data. All the six best performing log parsers have a drop in accuracy or show obvious fluctuations as the log volume increases. The log parsers, except IPLoM, are relatively stable on HDFS data, achieving an accuracy over $80\%$. Drain and AEL also show relatively stable accuracy on BGL data. However, on Android data, all the parsers have a large degradation on accuracy, because Android logs have quite a large number of event templates and are more complex to parse. Compared to other log parsers, Drain achieves relatively stable accuracy and shows its robustness when changing volumes of logs.

\subsection{Efficiency of Log Parsers}
Efficiency is an important aspect of log parsers to consider in order to handle log data in large scale. To measure the efficiency of a log parser, we record the running time it needs to finish the entire parsing process. Similar to the setting of the previous experiment, we evaluate six log parsers on three log datasets. 

The results are presented in Figure \ref{fig:time_datasize}. It is obvious that the parsing time increases with the raising of log size on all the three datasets. Drain and IPLoM have better efficiency, which scales linearly with the log size. Both methods can finish parsing 1GB of logs within tens of minutes. AEL also performs well except on large BGL data. It is because AEL needs to compare with every log message in a bin, yet BGL has a large bin size when the dataset is large. Other log parsers do not scale well with the volume of logs. Especially, LenMa and MoLFI cannot even finish parsing 1GB of BGL data or Android data within 6 hours. The efficiency of a log parser also depends on the type of logs. When the log data is simple and has a limited number of event templates, log parsing is often an efficient process. For instance, HDFS logs contain only 30 event templates, thus all the log parsers can process 1GB of data within an hour. However, the parsing process would become slow for logs with a large number of event templates (e.g., Android).

\section{Industrial Deployment}\label{sec:case_study}
In this section, we share our experiences of deploying automated log parsing in production at Huawei. System X (anonymized name) is one of the popular products of Huawei. Logs are collected during the whole product lifecycle, from development, testing, beta testing, to online monitoring. They are used as a main data source to failure diagnosis, performance optimization, user profiling, resource allocation, and some other tasks for improving product quality. When the system is still in a small scale, many of these analysis tasks are able to be performed manually. However, after a rapid growth in recent years, System X nowadays produces over terabytes of log data daily. It becomes impractical for engineers to manually inspect logs for diagnostic information, which requires not only non-trivial efforts but also deep knowledge of the logs. In many cases, event statistics and correlations are valuable hints to help engineers make informed decisions. 

To reduce the efforts of engineers, a platform (called LogKit) has been built to automate the log analysis process, including log search, rule-based diagnosis, and dashboard reporting of event statistics and correlations. A key feature of this platform is to parse logs into structured data. At first, log parsing was done in an ad-hoc way by writing regular expressions to match the events of interest. However, the parsing rules become unmanageable quickly. First, existing parsing rules cannot cover all types of logs, since it is time-consuming to write the parsing rules one by one. Second, System X is evolving quickly, leading to frequent changes of log structures. Maintenance of such a rule base for log parsing has become a new pain point. As a result, automated log parsing is a high demand. 

\textbf{Success stories}. With close collaboration with the product team, we have successfully deployed automated log parsing in production. After detailed comparisons of different log parsers as described in Section~\ref{sec:experiment}, we choose Drain because of its superiority in accuracy, robustness, and efficiency. In addition, by taking advantage of the characteristics of the logs of System X, we have optimized the Drain approach from the following aspects. \textit{1) Preprocessing}. The logs of System X have over ten thousand event templates as well as a wide range of parameters. As we have done in~\cite{He_dsn_2016}, we apply a simple yet effective preprocessing step to filter common parameters, such as IP, package name, number, and file path. This greatly simplifies the problem for subsequent parsing. Especially, some of the preprocessing scripts are extracted from the original parsing rule base, which is already available. \textit{2) Deduplication}. Many log messages comprise only constant string, with no parameters inside (e.g., ``VM terminated."). Recurrences of these log messages result in a large number of duplicate messages in logs. Meanwhile, the preprocessing step produce a lot of duplicate log messages as well (e.g., ``Connected to $<$IP$>$"), in which common parameters have been removed. We perform deduplication of these duplicate log messages to reduce the data size, which significantly improves the efficiency of log parsing. \textit{3) Partitioning}. The log message header contains two fields: verbosity level and component. In fact, log messages of different levels or components are always printed by different logging statements (e.g., DEBUG vs. INFO). Therefore, it is beneficial to partition log messages into different groups according to the level and component information. This naturally divides the original problem into independent subproblems. \textit{4) Parallelization}. The partitioning of logs can not only narrow down the search space of event templates, but also allow for parallelization. In particular, we extend Drain with Spark and naturally exploit the above log data partitioning for quick parallelization. By now, we have successfully run Drain in production for more than one year, which attains over 90\% accuracy in System X. We believe that the above optimizations are general and can be easily extended to other similar systems as well. 

\textbf{Potential improvements}. During the industrial deployment of Drain, we have observed some directions that need further improvements. \textit{1) State identification}. State variables are of significant importance in log analysis (e.g., ``DB connection ok" vs. ``DB connection error"). However, current log parsers cannot distinguish state values from other parameters. \textit{2) Dealing with log messages with variable lengths}. A single logging statement may produce log messages with variable lengths (e.g., when printing a list). Current log parsers are length-sensitive and fail to deal with such cases, thus resulting in degraded accuracy. \textit{3) Automated parameters tuning}. Most of current log parsers apply data-driven approaches to extracting event templates and some model parameters need to be tuned manually. It is desirable to develop a mechanism for automated parameters tuning. We call for research efforts to realize the above potential improvements, which would contribute to better adoption of automated log parsing.

\section{Related Work}\label{sec:relatedwork}
Log parsing is only a small part of the broad problem of log management. In this section, we review the related work from the aspects of log quality, log parsing, and log analysis.


\textbf{Log quality.}
The effectiveness of log analysis is directly determined by the quality of logs. To enhance log quality, recent studies have been focused on providing informative logging guidance or effective logging mechanisms during development. Yuan et al.~\cite{yuanICSE12} and Fu et al.~\cite{fu2014developers} report the logging practices in open-source and industrial systems, respectively. Zhu et al. \cite{zhu2015learning} propose LogAdvisor, a classification-based method to make logging suggestions on where to log. Zhao et al. \cite{Zhao_sosp_log20} further provide an entropy metric to determine logging points with maximal coverage of a control flow. Yuan et al.~\cite{yuanASPLOS11} design LogEnhancer to enhance existing logging statements with informative variables. Recently, He et al.~\cite{he2018ASE} have conducted an empirical study on the natural language descriptions of logging statements. Ding et al. \cite{dingATC15} provide a cost-effective way for dynamic logging with limited overhead. 
 
\textbf{Log parsing.}
Log parsing has been widely studied in recent years, which can be categorized into rule-based, source code-based, and data-driven parsing. Most current log management tools support rule-based parsing (e.g.,~\cite{loggly_autoparsing,logz_autoparsing}). Some studies~\cite{Xu_sosp_2009,Nagappan_issre_2009} make use of static analysis techniques for source code-based parsing. Data-driven log parsing approaches are the main focus of this paper, most of which have been summarized in Section~\ref{sec:background}. More recently, He et al.~\cite{HeZHLL18} have studied large-scale log parsing through the parallelization on Spark. Thaler et al.~\cite{Thaler_isdfs17} model textual log messages with deep neural networks. Gao et al.~\cite{GaoHP18} apply an optimization algorithm to discover multi-line structures from logs.

\textbf{Log analysis.}
Log analysis is a research area that has been studied for decades due to its practical importance. There are an abundance of techniques and applications of log analysis. Typical applications include anomaly detection \cite{Xu_sosp_2009, fu2009execution,He_issre_2016,HeLLZLZ18}, problem diagnosis \cite{Yuan_ASPLOSP10,Nagaraj_nsdi_2012}, runtime verification \cite{wshang13}, performance modeling \cite{Chow_osdi_2014}, etc. To address the challenges involved in log analysis, many data analytics techniques have been developed. For example, Xu et al. \cite{Xu_sosp_2009} apply the principle component analysis (PCA) to identify anomaly issues. Du et al.~\cite{Du_ccs_2017} investigate the use of deep learning to model event sequences. Lin et al.~\cite{linqwICSE16} develop a clustering algorithm to group similar issues. Our work on log parsing serves as the basis to perform such analysis and can greatly reduce the efforts for the subsequent log analysis process.


\section{Conclusion}\label{sec:conclusion}
Log parsing plays an important role in system maintenance, because it serves as the the first step towards automated log analysis. In recent years, many research efforts have been devoted towards automated log parsing. 
However, there is a lack of publicly available log parsing tools and benchmark datasets. In this paper, we implement a total of 13 log parsing methods and evaluate them on 16 log datasets from different types of software systems. We have opened source our toolkit and released the benchmark datasets to researchers and practice for easy reuse. Moreover, we share the success stories and experiences when deploying automated log parsing at Huawei. We hope our work, together with the released tools and benchmarks, could facilitate more research on log analysis.

\section*{Acknowledgment}
The work described in this paper was supported by the National Key R\&D Program of China (2018YFB1004804), the National Natural Science Foundation of China (61722214, 61502401, 61332010), the Program for Guangdong Introducing Innovative and Entrepreneurial Teams (2016ZT06D211), the Research Grants Council of the Hong Kong Special Administrative Region, China (No. CUHK 14210717 of the General Research Fund). Zibin Zheng is the corresponding author.

\bibliographystyle{IEEEtran}
\bibliography{icse} 

\balance

\end{document}